\documentclass[12pt,preprint]{aastex}

\newcommand{\etal}{et~al.\ }
\newcommand{\CaIIdblt}{{\rm Ca}\kern 0.1em{\sc ii}~$\lambda\lambda 3934, 3969$}
\newcommand{\CIVdblt}{{\rm C}\kern 0.1em{\sc iv}~$\lambda\lambda 1548, 1550$}
\newcommand{\MgIIdblt}{{\rm Mg}\kern 0.1em{\sc ii}~$\lambda\lambda 2796, 2803$}
\newcommand{\NVdblt}{{\rm N}\kern 0.1em{\sc v}~$\lambda\lambda 1238, 1242$}
\newcommand{\SVIdblt}{{\rm S}\kern 0.1em{\sc vi}~$\lambda\lambda 933, 944$}
\newcommand{\OVIdblt}{{\rm O}\kern 0.1em{\sc vi}~$\lambda\lambda 1031, 1037$}
\newcommand{\SiIIdblt}{{\rm Si}\kern 0.1em{\sc ii}~$\lambda\lambda 1190, 1193$}
\newcommand{\SiIVdblt}{{\rm Si}\kern 0.1em{\sc iv}~$\lambda\lambda 1393, 1402$}
\newcommand{\PV}{\hbox{{\rm P}\kern 0.1em{\sc v}}}
\newcommand{\AlI}{\hbox{{\rm Al}\kern 0.1em{\sc i}}}
\newcommand{\AlII}{\hbox{{\rm Al}\kern 0.1em{\sc ii}}}
\newcommand{\AlIII}{{\hbox{\rm Al}\kern 0.1em{\sc iii}}}
\newcommand{\CaII}{\hbox{{\rm Ca}\kern 0.1em{\sc ii}}}
\newcommand{\CII}{\hbox{{\rm C}\kern 0.1em{\sc ii}}}
\newcommand{\CIIe}{\hbox{{\rm C$^{\ast}$}\kern 0.1em{\sc ii}}}
\newcommand{\CIII}{\hbox{{\rm C}\kern 0.1em{\sc iii}}}
\newcommand{\CIV}{\hbox{{\rm C}\kern 0.1em{\sc iv}}}
\newcommand{\CV}{\hbox{{\rm C}\kern 0.1em{\sc v}}}
\newcommand{\HI}{\hbox{{\rm H}\kern 0.1em{\sc i}}}
\newcommand{\HII}{\hbox{{\rm H}\kern 0.1em{\sc ii}}}
\newcommand{\Ha}{\hbox{{\rm H}\kern 0.1em$\alpha$}}
\newcommand{\Lya}{\hbox{{\rm Ly}\kern 0.1em$\alpha$}}
\newcommand{\Lyb}{\hbox{{\rm Ly}\kern 0.1em$\beta$}}
\newcommand{\Lyg}{\hbox{{\rm Ly}\kern 0.1em$\gamma$}}
\newcommand{\Lyd}{\hbox{{\rm Ly}\kern 0.1em$\delta$}}
\newcommand{\Lye}{\hbox{{\rm Ly}\kern 0.1em$\epsilon$}}
\newcommand{\Lyphi}{\hbox{{\rm Ly}\kern 0.1em$\phi$}}
\newcommand{\Lyfive}{\hbox{{\rm Ly}\kern 0.1em$5$}}
\newcommand{\Lysix}{\hbox{{\rm Ly}\kern 0.1em$6$}}
\newcommand{\Lyseven}{\hbox{{\rm Ly}\kern 0.1em$7$}}
\newcommand{\Lyeight}{\hbox{{\rm Ly}\kern 0.1em$8$}}
\newcommand{\Lynine}{\hbox{{\rm Ly}\kern 0.1em$9$}}
\newcommand{\Lyten}{\hbox{{\rm Ly}\kern 0.1em$10$}}
\newcommand{\Lyeleven}{\hbox{{\rm Ly}\kern 0.1em$11$}}
\newcommand{\HeI}{\hbox{{\rm He}\kern 0.1em{\sc i}}}
\newcommand{\HeII}{\hbox{{\rm He}\kern 0.1em{\sc ii}}}
\newcommand{\FeI}{\hbox{{\rm Fe}\kern 0.1em{\sc i}}}
\newcommand{\FeII}{\hbox{{\rm Fe}\kern 0.1em{\sc ii}}}
\newcommand{\FeIII}{\hbox{{\rm Fe}\kern 0.1em{\sc iii}}}
\newcommand{\MnII}{\hbox{{\rm Mn}\kern 0.1em{\sc ii}}}
\newcommand{\MgI}{\hbox{{\rm Mg}\kern 0.1em{\sc i}}}
\newcommand{\MgII}{\hbox{{\rm Mg}\kern 0.1em{\sc ii}}}
\newcommand{\MgIII}{\hbox{{\rm Mg}\kern 0.1em{\sc iii}}}
\newcommand{\NiI}{\hbox{{\rm Ni}\kern 0.1em{\sc i}}}
\newcommand{\NI}{\hbox{{\rm N}\kern 0.1em{\sc i}}}
\newcommand{\NII}{\hbox{{\rm N}\kern 0.1em{\sc ii}}}
\newcommand{\NIII}{\hbox{{\rm N}\kern 0.1em{\sc iii}}}
\newcommand{\NV}{\hbox{{\rm N}\kern 0.1em{\sc v}}}
\newcommand{\OVI}{\hbox{{\rm O}\kern 0.1em{\sc vi}}}
\newcommand{\OI}{\hbox{{\rm O}\kern 0.1em{\sc i}}}
\newcommand{\OII}{\hbox{[{\rm O}\kern 0.1em{\sc ii}]}}
\newcommand{\OIV}{\hbox{{\rm O}\kern 0.1em{\sc iv}]}}
\newcommand{\SI}{{\rm S}\kern 0.1em{\sc i}}
\newcommand{\SIV}{{\rm S}\kern 0.1em{\sc iv}}
\newcommand{\SVI}{{\rm S}\kern 0.1em{\sc vi}}
\newcommand{\SiI}{\hbox{{\rm Si}\kern 0.1em{\sc i}}}
\newcommand{\SiII}{\hbox{{\rm Si}\kern 0.1em{\sc ii}}}
\newcommand{\SiIII}{\hbox{{\rm Si}\kern 0.1em{\sc iii}}}
\newcommand{\SiIV}{\hbox{{\rm Si}\kern 0.1em{\sc iv}}}
\newcommand{\SII}{\hbox{{\rm S}\kern 0.1em{\sc ii}}}
\newcommand{\SIII}{\hbox{{\rm S}\kern 0.1em{\sc iii}}}
\newcommand{\NaI}{\hbox{{\rm Na}\kern 0.1em{\sc i}}}
\newcommand{\TiII}{\hbox{{\rm Ti}\kern 0.1em{\sc ii}}}
\newcommand{\CrII}{\hbox{{\rm Cr}\kern 0.1em{\sc ii}}}
\newcommand{\ZnII}{\hbox{{\rm Zn}\kern 0.1em{\sc ii}}}
\newcommand{\kms}{\hbox{km~s$^{-1}$}}
\newcommand{\cmsq}{\hbox{cm$^{-2}$}}
\newcommand{\cc}{\hbox{cm$^{-3}$}}

\shorttitle{THE $z=1.39$ DLA TOWARD Q~$0957+561$ A,B}
\shortauthors{CHURCHILL ET AL.}

\begin{document}

\title{The Spatial, Ionization, and Kinematic Conditions of the
$\lowercase{z}=1.39$ Damped {\Lya} Absorber in Q$0957+561$
A,B\altaffilmark{1}}

\author{Christopher W. Churchill\altaffilmark{2}, 
Richard R. Mellon\altaffilmark{3}, 
Jane C. Charlton\altaffilmark{4}}
\affil{Department of Astronomy and Astrophysics \\
       The Pennsylvania State University \\
       University Park, PA 16802 \\
       {\it cwc, rmellon, charlton@astro.psu.edu}} 

\and

\author{Steven S. Vogt\altaffilmark{2}}
\affil{UCO/Lick Observatories \\
       Board of Studies in Astronomy and Astrophysics \\
       University of California, Santa Cruz, CA 96054 \\
       {\it vogt@ucolick.org}}

\altaffiltext{1}{Based  in  part   on  observations  obtained  at  the
W.~M. Keck Observatory, which  is operated as a scientific partnership
among Caltech, the University of California, and NASA. The Observatory
was made possible by the  generous financial support of the W.~M. Keck
Foundation.}
\altaffiltext{2}{Visiting Astronomer, the W.~M. Keck Observatory}
\altaffiltext{3}{Presently at the Department of Astronomy, University
of Virginia} 
\altaffiltext{4}{Center for Gravitational Physics and Geometry}

\begin{abstract}

We examined the kinematics, ionization conditions, and physical size
of the absorption clouds in a $z=1.3911$ damped {\Lya} absorber (DLA)
in the double image lensed quasar Q~$0957+561$ A,B (separation
$135h_{75}^{-1}$~pc at the absorber redshift).  Using HIRES/Keck
spectra (FWHM$~\simeq 6.6$~{\kms}), we studied the {\MgIIdblt}
doublet, {\FeII} multiplet, and {\MgI} $\lambda 2853$ transition in
absorption.  Based upon the {\FeII} profiles (the {\MgII} suffers from
saturation), we defined six ``clouds'' in the system of sightline A
and seven clouds in system of sightline B.  An examination of the
$N(v)$ profiles, using the apparent optical depth method, reveals no
clear physical connection between the clouds in A and those in B.  The
observed column density ratios of all clouds is $\log
N({\MgI})/N({\FeII}) \simeq -2$ across the full $\sim 300$~{\kms}
velocity range in both systems and also spatially (in both
sightlines).  This is a remarkable uniformity not seen in Lyman limit
systems.  The uniformity of the cloud properties suggests that the
multiple clouds are not part of a ``halo''.  Based upon
photoionization modeling, using the $N({\MgI})/N({\FeII})$ ratio in
each cloud, we constrain the ionization parameters in the range $-6.2
\leq \log U \leq -5.1$, where the range brackets known abundance ratio
and dust depletion patterns.  The inferred cloud properties are
densities of $2 \leq n_{\rm H} \leq 20$~{\cc}, and line of sight sizes
of $1 \leq D \leq 25$~pc.  The masses of the clouds in system A are
$10 \leq M/M_{\odot} \leq 1000$ and in system B are $1 \leq
M/M_{\odot} \leq 60$ for spherical clouds.  For planar clouds, the
upper limits are $400h_{75}^{-2}$~M$_{\odot}$ and
$160h_{75}^{-2}$~M$_{\odot}$ for A and B, respectively.  We favor a
model of the absorber in which the DLA region itself is a single cloud
in this complex, which could be a parcel of gas in a galactic ISM.  We
cannot discern if the {\HI} in this DLA cloud is in a single, cold
phase or in cold+warm phases.  A spherical cloud of $\sim 10$~pc would
be limited to one of the sightlines (A) and imply a covering factor
less than 0.1 for the DLA complex.  We infer that the DLA cloud
properties are consistent with those of lower density, cold clouds in
the Galactic interstellar medium.

\end{abstract}

\keywords{
quasars: absorption lines --- quasars: individual (Q0957+561 A,B) ---
galaxies: ISM 
}

\section{Introduction}
\label{sec:intro}

Damped {\Lya} absorbers (DLAs) are well appreciated as useful
astronomical laboratories for measuring metal enrichment,
nucleosynthetic processing, and dust evolution from $z=4$ to the
present epoch
\citep[e.g.][]{pettini94,lu96,pettini97,vladilo98,pettini99,pw00,prochaska02,ledoux02}.
Their redshift number density evolution \citep{wolfe95,lisa96,rt00}
places constraints on $\Omega ({\HI},z)$, and probes the {\HI} mass
function for column densities greater than $N({\HI}) = 2\times
10^{20}$~{\cmsq}.  The gas kinematics measured from low--ionization
metal absorption lines in DLAs provides valuable data upon which
hypotheses for their dynamical formation, evolution, and physical
nature can be based.  To date, these competing hypotheses include
thick rotating disks \citep{pw97,pw98}, merging cold dark matter halos
\citep{haehnelt98,mcdonald99}, and supernovae winds \citep[][however,
see Bond et~al.\ 2001]{nulsen98,shaye01}.

The fact that DLAs provide leverage on questions relating to
cosmological chemical evolution, nucleosynthetic and dust evolution,
galaxy and/or hierarchical clustering evolution, and global star
formation evolution makes them one of the most useful, and therefore,
important astronomical objects of study.  Of the great deal that has
been learned about DLAs, one property remains elusive; there is little
direct observational evidence from which the characteristic size of
the DLA ``region'' itself can be deduced.  By DLA region, we mean the
parcel of gas that is associated with the very large {\HI} column density
greater than $N({\HI}) = 2 \times 10^{20}$~{\cmsq}.  From the size
estimates, one could then deduce the covering factor of DLAs within
their host objects, as well as their densities, and most importantly,
their masses (though this quantity is geometry dependent).

Associating DLAs with the normal, bright galaxy population and
assuming a unity covering factor, \citet{steidel93} deduced a
characteristic size\footnote{Throughout this paper we use $h_{75} =
H_{0}/75$ and $q_{0}=0$.} of $\sim 20h^{-1}_{75}$~kpc.  However, it is
now known that DLAs are associated with galaxies having a wide variety
of morphological types, including $0.1~L^{\ast}$ galaxies and low
surface brightness (LSB) galaxies \citep{lebrun97,rt98,bowen01}.
Indeed, the luminous host of many DLAs remain undetected even after
dedicated searches to find their diffuse {\Lya} emission
\citep[e.g.,][]{deharveng90,lowenthal95}, their optical continuum
emission \citep[e.g.,][]{steidel97,kulkarni00,kulkarni01} and their
{\Ha} emission
\citep[e.g.,][]{bunker99,bouche00,kulkarni00,kulkarni01}.

In the absence of a well--defined luminosity function for the luminous
hosts of DLAs, and without knowledge of the covering factor of the DLA
region in these objects, we remain ignorant of a characteristic DLA
size.  It is safe to say, however, that estimates incorporating low
surface brightness galaxies in the luminosity function of known DLA
hosts would drive down the characteristic size.  Scaling the Steidel
estimate by the ratio of $\Phi _{\ast}$ for the luminosity functions
of normal bright galaxies \citep{lilly95,ellis96,lin99} and LSB
galaxies \cite[e.g.,][]{dalcanton97}, we find an approximate
characteristic size of $(0.033/(0.033+0.08)\cdot 20h^{-1}_{75} \sim
6h_{75}^{-1}$~kpc.
 
The data directly constraining the physical sizes of DLAs are few.  In
the lensed quasar HE~$1104-1805$, \citet{smette95} placed limits of
$5$--$17h^{-1}_{75}$~kpc on a DLA in one of the images.  A more
constraining result was obtained by \citet{michal97}, who used the
lensed quasar Q~$0957+561$ A,B to show that the physical extent of a
DLA at $z=1.3911$ is less than $135h_{75}^{-1}$~pc.  On the other
hand, \citet{petitjean00} suggest a lower limit of $200h_{75}^{-1}$~pc
for a probable DLA in the lensed quasar APM $08279+5255$ (based upon
partial covering arguments).  Using VLBA observations of the slightly
extended radio quasar B~$0738+313$, \citet{lane00} find $\log N({\HI})
= 21.2$~{\cmsq} in 21--cm absorption at two locations separated by
$20h^{-1}_{75}$ pc [we note, however, that 21--cm emission line
techniques may not be selecting objects that are one and the same as
the DLAs selected using quasar absorption line techniques
\citep[see][]{rt00,cwc01}].

In addition to the great insights garnished from low-- to
moderate--resolution spectroscopic studies of multiply imaged quasars,
high resolution spectra provide constraints on the kinematics of
metal--line systems at the few kilometers per second level
\citep[e.g.,][]{rauch99-i,lopez99,petitjean00,rauch02-iv}.  This means
the gaseous structure can be studied component by component, directly
providing their sightline to sightline velocity shears, $\Delta
v/\Delta r$, where $r$ is the physical separation of the sightlines at
the absorber.  Such measurements allow insights into the ionization
structures, masses, density gradients, and energy input budgets
\citep[e.g.,][]{rauch99-i,rauch01-iii,dodorico02}.  It is important to
realize that these insights are derived based upon models of the gas
dynamics and geometry.  These models, however, are strongly founded
upon observations of the gaseous components in the Milky Way, LMC,
SMC, and other local galaxies.

The reasonably bright gravitationally lensed quasar Q~$0957+561$ A,B
at $z_{em} = 1.41$ with separation 6{\arcsec} provides an excellent
opportunity to study high--resolution spectra of two sightlines
through a $z>1$ DLA
\citep[e.g.,][]{walsh79,wills80,young80,young81a,young81b,turnshek93,michal97,zuo97,pettini99}.
The $z=1.3911$ absorber has a neutral hydrogen column density of
$N({\HI}) = 1.9\pm 0.3\times 10^{20}$~{\cmsq} in the A spectrum and
$N({\HI})=8\pm2 \times 10^{19}$~{\cmsq} in the B spectrum
\citep{zuo97}.  This suggests that the DLA region is concentrated in
front of the A image.  At the redshift of the absorber, the physical
separation of the A and B sightlines is $135h_{75}^{-1}$~pc
\citep{smette92}.  \citet{zuo97} found a differential reddening in
spectrum A as compared to spectrum B and deduce a dust to gas ratio of
$\sim 0.6$ in the DLA.  Imaging of the quasar field has been reported
by \citet[][ground based]{dalhe94} and
\citet[][WFPC2/HST]{bernstein97}.  Despite these efforts, the luminous
object associated with the $z=1.3911$ absorber remains unidentified.

The most extensive study of the DLA to date, using FOS/{\it HST}
spectra ($R=1300$), was presented by \citet{michal97}.  In general,
they found ``the differences in the damped Lyman series absorption in
the lensed components are the only significant spectral
characteristics that distinguishes the far--ultraviolet spectra of
Q~$0957+561$ A,B.''  Assuming a spherical geometry and a neutral
hydrogen fraction of $N({\HI})/N({\rm H}) = 0.05$ \citep{chartas95},
the authors estimate the DLA mass to be $2.6\times 10^{6}$~M$_
{\odot}$, which they interpreted to be consistent with a giant
molecular cloud or a small condensation (spiral arm) within a galactic
disk (viewed pole on).  Because the metal lines are tightly correlated
for the two sightlines, \citet{michal97} suggested that the large
metal--line absorption arises in a galactic ``halo''.

In this paper, we present $R=45,000$ HIRES/Keck spectra of the
Q~$0957+561$ A,B.  We focus on the {\MgIIdblt} doublet, the
{\MgI}~$\lambda 2853$ transition, and several {\FeII} transitions
arising in the DLA at $z=1.3911$.  In \S~\ref{sec:data}, we present
the data, and briefly describe the data reduction and analysis.  We
present the observed kinematic and spatial properties of the absorber
column densities in \S~\ref{sec:cols}.  Assuming photoionization
equilibrium, we model the clouds and present their densities, sizes,
and masses in \S~\ref{sec:results}.  We briefly discuss these
properties in \S~\ref{sec:discuss} and provide concluding remarks in
\S~\ref{sec:conclusions}.

\section{The Data}
\label{sec:data}

The lensed quasar Q~$0957+561$ A,B was observed on the nights of 1994
November 26--27 with the HIRES instrument \citep{vogt94} on the
Keck--I telescope.  HIRES was configured in first order using decker
C1 (7{\arcsec} in the spatial 0.861{\arcsec} in the dispersion
directions, respectively).  The resolution of $R=45,000$ corresponds
to a velocity resolution of ${\rm FWHM} = 6.6$~{\kms}.  The resulting
observed wavelength range is 4600 to 6900~{\AA}.  As such, many
different transitions were covered; however, we detected only the
{\MgIIdblt} doublet, the {\FeII} $\lambda$2344, 2374, 2383, 2587, \&
2600 multiplet, and the {\MgI} $\lambda 2853$ transition.

The quasar images A and B were observed individually.  Three 1500
second exposures were obtained for each image. The resulting spectra
have signal--to--noise ratios of $\simeq 30$ per resolution element.
The individual spectra were reduced and calibrated using the
IRAF\footnote{IRAF is distributed by the National Optical Astronomy
Observatories, which are operated by AURA, Inc., under contract to the
NSF.} {\it Apextract\/} package for echelle data.  The wavelength
scale is vacuum and has been corrected to the heliocentric velocity.
Continuum normalization was performed as described in
\citet{sembach92} and \citet{thesis}.

\begin{figure*}[p]
\plotone{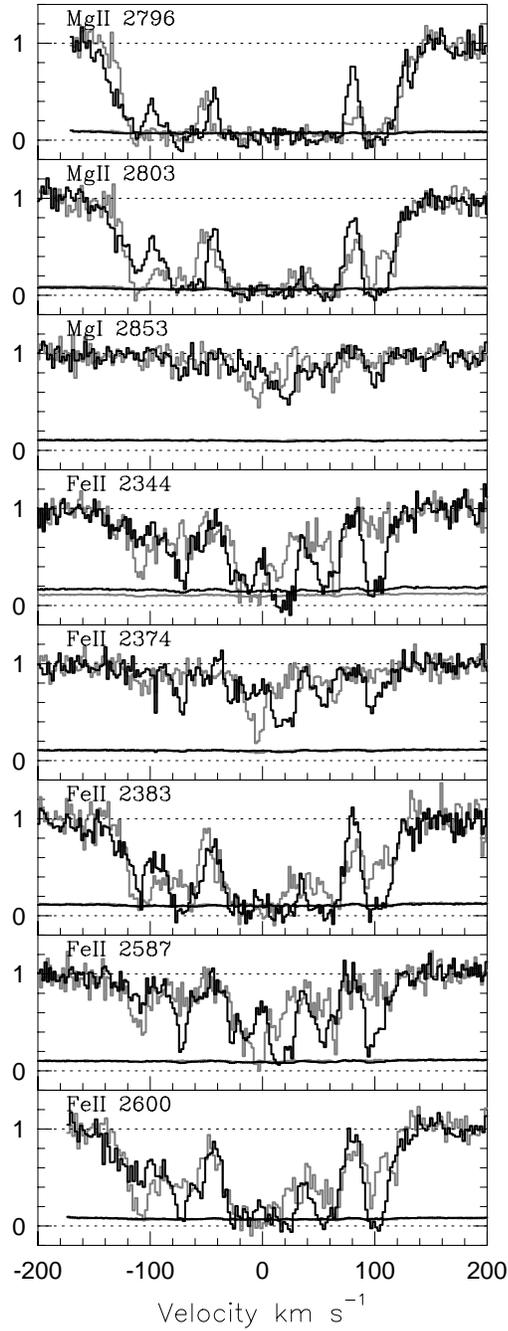}
\figurenum{1}
\caption
{Normalized HIRES/Keck spectra of the {\MgII}, {\MgI}, and {\FeII} 
absorption profiles presented in the system rest--frame velocity.
A spectra are black and B spectra are grey.  The velocity
zero point is set to $z = 1.390861$, which is the optical depth mean
of the {\MgII} $\lambda 2796$ transition in the A spectrum.
\label{fig:spectra}}
\end{figure*}

In Figure~\ref{fig:spectra}, we present the observed {\MgII}, {\MgI},
and {\FeII} absorption profiles for the $z=1.391$ absorber as a
function of rest--frame velocity, where the A spectra (black) and B
spectra (grey) are over plotted on the same velocity scale.  The
velocity zero point is set to $z=1.390861$, which corresponds to the
optical depth mean of the {\MgII} $\lambda 2796$ profile for the
system in sightline A \citep[see Appendix A.1 of][]{cv01}.

\section{Column Densities and Kinematics}
\label{sec:cols}

We modeled the data using Voigt profile (VP) decomposition.  The
fitting was performed with the code MINFIT \citep{thesis,cvc02}.  The
VP models provide the number of clouds and their velocities, column
densities, and Doppler parameters.  Because VP analysis is non--unique
(especially in highly saturated profiles), and because it is sensitive
to the signal--to--noise ratio and resolution of the spectra
\citep{thesis,cvc02}, we use the VP parameters as a secondary means of
studying the column densities and kinematics.  Instead, we use the
apparent optical depth method \citep{savage91} as the primary means of
studying the relationship between cloud velocities and column
densities.

\subsection{Apparent Optical Depth Method}

Apparent column densities per unit velocity, $N_{a}(v)$
[atoms~{\cmsq}/({\kms})], were measured for each transition using the
formalism described by \citet{savage91}.  These column density spectra
and their uncertainty spectra were then linearized to a common
velocity binning of 2.23~{\kms} using flux conservation. From these
linearized data, an optimal column density was computed for each
species in each velocity bin.  The algorithm employed for computing
optimal column densities has been described in Appendix A.5 of
\citet{cv01}.

In brief, for adoption of the optimal $N_{a}(v)$ for an ion we 
employ one of three possibilities: (1)~All transitions of an ion exhibit some
saturation between $v_{b}$ and $v_{r}$, so that $N_{a}(v)$ is a lower
limit.  Often, the transition with the smallest $f\lambda$ provides
the best constraints on the lower limit; (2)~All but one transition of
an ion exhibits saturation, in which case, the adopted column density
is taken from the unsaturated transition; and (3)~All or more than one
transition of an ion are unsaturated, providing multiple independent
measurements.  The optimal $N_{a}(v)$ is computed from the weighted mean
in each velocity bin.  

The {\MgII} transitions in both systems exhibited unresolved
saturation across the majority of the profiles and therefore provide
only upper limits on $N_{a}(v)$ at most velocities.  Though the stronger
{\FeII} transitions ($\lambda 2383$ and $\lambda 2600$) exhibited
saturation over the velocity intervals $(-30,+30)$, $(+50,+70)$, and
$(+90,+110)$, the weakest transition ($\lambda 2374$), and to a lesser
extent the $\lambda 2587$ transition, provide a robust $N_{a}(v)$ across
these intervals.  {\MgI} has no velocity regions with unresolved
saturation.

\begin{figure*}[p]
\plotone{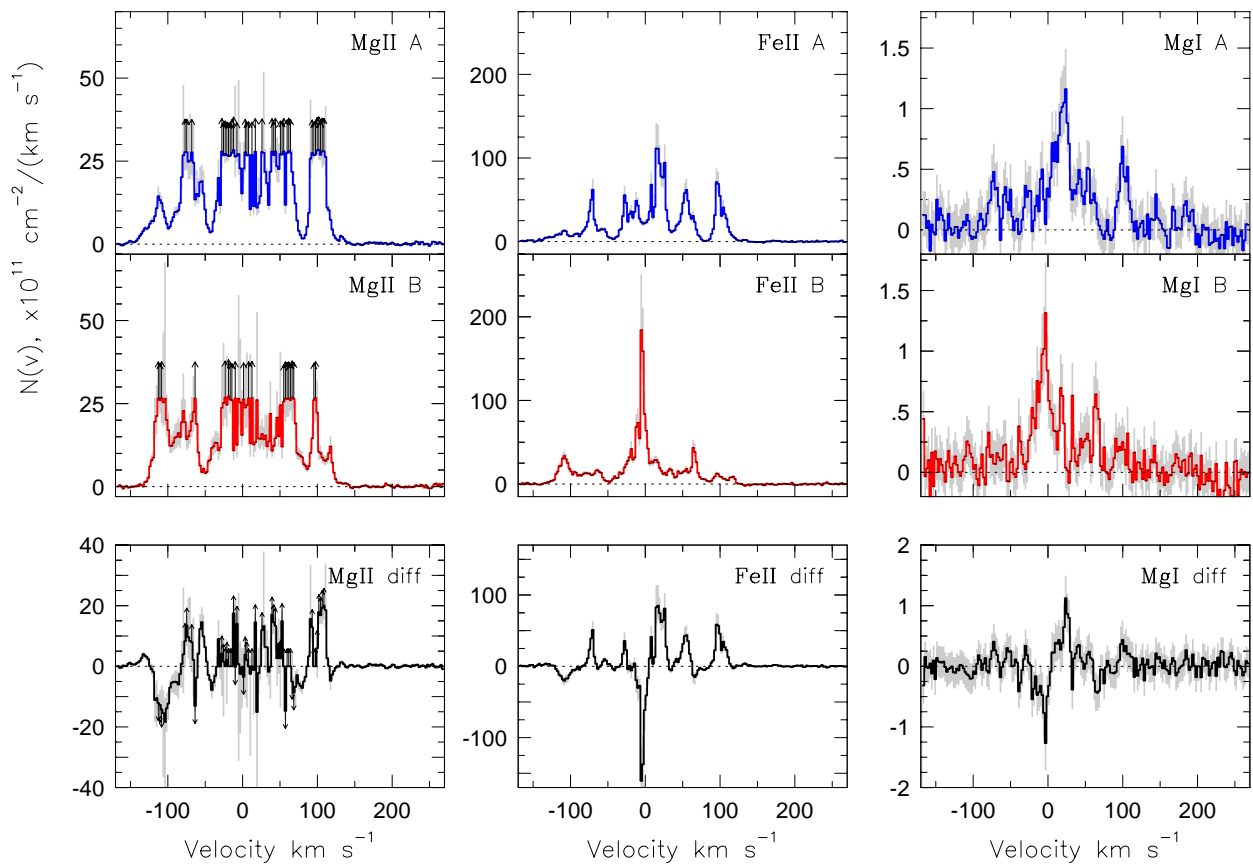}
\figurenum{2}
\caption
{The optimal apparent column densities profiles, $N(v)$, for {\MgII}
(left), {\FeII} (center), and {\MgI} (right) presented in the system
rest--frame velocity.  The grey shading is the $1~\sigma$
uncertainties in the $N(v)$ profiles.  The lower panels show the
differences A$-$B.
\label{fig:aods}}
\end{figure*}

In Figure~\ref{fig:aods}, we present $N(v)$, the optimal apparent
column densities, for {\MgII} (left), {\FeII} (center), and {\MgI}
(right) as a function of rest--frame velocity\footnote{From this point
in the text, we drop the subscript ``$a$'' designating ``apparent'' in
the AOD column densities.}.  The upper panels show $N(v)$ for systems
of sightlines A and B (hereafter, systems A and B) as labeled and the
lower panels show the column density difference, $\Delta N(v) =
N_{A}(v) - N_{B}(v)$.  Limits are represented by arrows and $1~\sigma$
uncertainties are shown as grey shading.

A prominent feature of the data in Figure~\ref{fig:aods}, is the
$N(v)$ peak for {\FeII} and {\MgI} in system A at $v\simeq
+20$~{\kms}.  A second, even more prominent $N(v)$ peak is at $v\simeq
-10$~{\kms} in system B.  Interestingly, for {\FeII}, the difference
profile, $\Delta N(v)$, reveals positive (A$>$B) and negative (A$<$B)
peaks with a quasi--periodicity of $\simeq 40$~{\kms}.  In some cases
these $\Delta N(v)$ peaks are due to different strengths of $N(v)$
peaks aligned in velocity in both systems.  In other cases the $\Delta
N(v)$ peaks are due to the presence of an $N(v)$ peak in the one
system and the lack of an $N(v)$ peak in the other at the same
velocity.

\subsection{Velocity Correlations}
\label{sec:xcorr}

A striking feature of the data is the velocity alignment of an {\it
absence\/} of absorption at $v \simeq + 80$~{\kms}, as can be clearly
seen in the {\MgII} and strongest {\FeII} profiles in
Figure~\ref{fig:spectra}.  This may indicate that the parcel of gas
giving rise to absorption at $v \simeq +100$~{\kms} is a separate {\it
physical\/} entity from that giving rise to the lower velocity
absorption.  If so, this gas has little to no velocity sheer across
the sightlines.  At $ v \simeq -50$~{\kms} and at $v \simeq
-100$~{\kms} there are similar, yet less pronounced, profile inversions.
In this case, there is a small difference of $\sim 20$~{\kms}, which
translates to a velocity sheer of $0.15 h_{75}$~{\kms}~pc$^{-1}$.
Again, this could indicate that the higher (negative) velocity gas is
physically distinct from the lower velocity gas.

\begin{figure*}[p]
\plotone{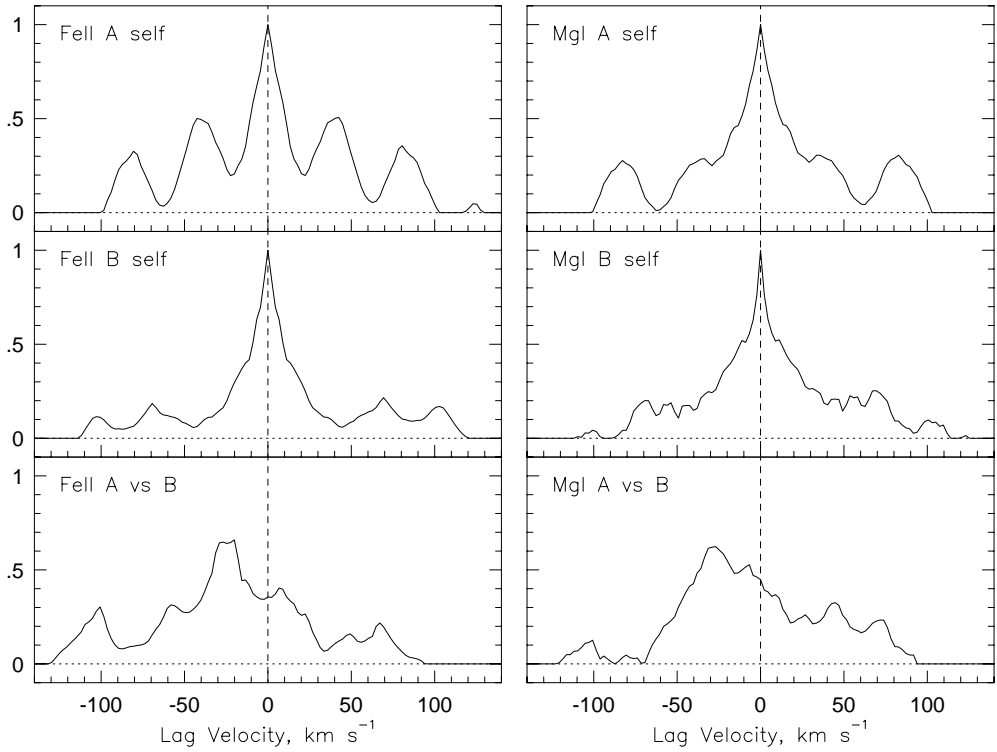}
\figurenum{3}
\caption
{The cross--correlation function, $\xi (\Delta v)$, for the {\FeII}
and {\MgI} $N(v)$ profiles.  Left hand panels are {\FeII} with the
self--correlation of A (top), self--correlation of B (middle), and
cross--correlation of A with B (lower).  The right hand panels are
{\MgI}.
\label{fig:xcorr}}
\end{figure*}

Ultimately, it is difficult to extract physical information directly
from the flux values of such strong absorption lines, which exhibit
saturation over much of the velocity interval.  In order to further
study the cloud--by--cloud kinematic connections between systems A and
B, we ran a cross--correlation on the {\FeII} and {\MgI} $N(v)$
profiles.  The cross--correlation function is defined by
\begin{equation}
\xi (\Delta v) = \frac 
{
\sum \left[ N_{A}(v)-\left< N_{A} \right> \right] \cdot 
     \left[ N_{B}(v-\Delta v)-\left< N_{B} \right> \right] 
}
{ \sqrt{ 
  \sum \left[ N_{A}(v)-\left< N_{A} \right> \right] ^{2} \cdot 
  \sum \left[ N_{B}(v-\Delta v)-\left< N_{B} \right> \right] ^{2} 
  } 
} ,
\label{eq:xcorr}
\end{equation}
where $\Delta v$ is the lag velocity between the two systems.
Equation~\ref{eq:xcorr} is defined so that a perfect correlation is
$\xi = +1$ and no correlation is $ \xi = 0$ In Figure~\ref{fig:xcorr},
we present $\xi$ as a function of $\Delta v$.  The left hand panels
are {\FeII} and the right hand panels are {\MgI}.  The top (middle)
panels show the self--correlation function for system A (B).  By
definition, these functions are symmetric about the lag velocity
and have unity at zero lag velocity.

The $\xi$ for {\FeII} in system A shows a remarkable pattern; there is
a $40$~{\kms} periodicity in the clouds, as evident in the two peaks
in $\xi$ at $\left| \Delta v \right| \simeq 40$ and $80$~{\kms}.  There is
a lack of periodicity in the {\FeII} for system B; the
cross--correlation function reveals the velocity difference of the two
strongest clouds separated by $\Delta v \simeq 80$~{\kms}.  The clouds in
system A, in general, have large $N$, so it is quite clear that this
periodicity in system A is driving the shape of the $\Delta N(v)$
profile shown in Figure~\ref{fig:aods}.

The $\xi$ for {\MgI} shows no periodicity, but only the velocity
differences between the strongest components.  Note the ``noise'' in
$\xi$ for system B {\MgI} at $30 \leq \left| \Delta v \right| \leq
60$~{\kms}, which has a magnitude of $\delta \xi \simeq 0.1$; this
arises in the noisiest data and provides an estimate of the
significance level of the stronger peaks for all of the
cross--correlation functions.

In the lower panels of Figure~\ref{fig:xcorr}, we show the
cross--correlation functions for system A against system B.  The peaks
in these functions provide the velocity difference between the
strongest components in each system, which lies at $\Delta v \simeq
-27$~{\kms} for both {\FeII} and {\MgI}.  A peak of $\xi \simeq 0.7$
is significantly below unity (approximately $3~\sigma$, based upon the
above noise estimates) and quantifies the level at which the two
{\FeII} profiles do not resemble each other kinematically; $\xi$ is
dominated by the strongest components in complex profiles.  

Overall, this exercise reveals that there is no clear signal in the
cross--correlation function for similar kinematics in the system A and
system B profiles.  This indicates that the clouds are not clearly
traceable between the two sightlines.

\subsection{Integrated Column Densities}

In Table~\ref{tab:acds}, we present the integrated apparent column
densities, $N$, for the {\MgII}, {\FeII} and {\MgI} ions.  The $N$ were
computed for fixed velocity intervals (from $v_{1}$ to $v_{2}$) using
the data presented in Figure~\ref{fig:aods}.  The velocity intervals
were defined by local $N(v)$ minima in the {\FeII} spectra for system
A and B individually.  Six velocity intervals were found for system A
and seven were found for system B.  These intervals roughly represent
individual ``clouds'' giving rise to the complex absorption profiles.

As stated above, the {\MgII} transitions provide only upper limits for
most velocity intervals.  The exception is velocity interval 1 (or
``cloud number 1'') in system A.  Clouds 2, 3, 5, and 7 in systems B
are marginally saturated; there is at least one saturated pixel in
each of these clouds.  For this reason, we quote these particular
clouds in system B as upper limits.  However, unless there are high
column density clouds with $b \leq 2$~{\kms} at the location of these
saturated pixels, the quoted values for system B could be marginally
acceptable as measurements.  We choose to not invoke them as
measurements for our analysis.

\begin{figure*}[p]
\plotone{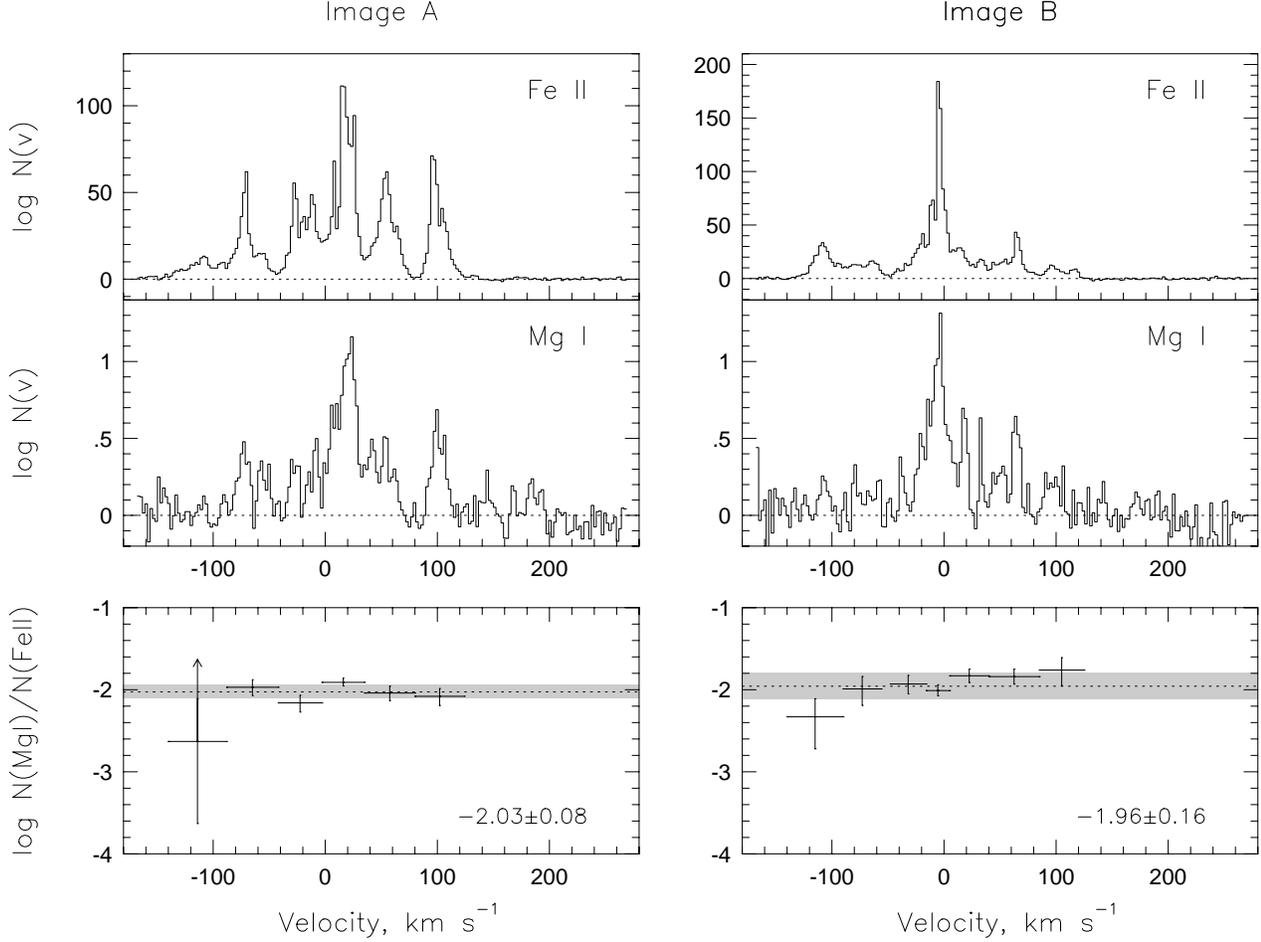}
\figurenum{4}
\caption
{The optimal apparent column densities profiles, $N(v)$, for {\FeII}
and {\MgI} presented in the system rest--frame velocity.  A
spectra are on the left and B spectra on the right.  The
bottom panels show the ratios of the integrated column densities for
each of the velocity bins corresponding to ``separate clouds'' in the
{\FeII} profiles.  The dotted horizontal line is the average of the
individual clouds and the grey shaded area is the standard deviation.
This average and standard deviation are written in the corner of the
respective panels.
\label{fig:rats}}
\end{figure*}

For both systems, we computed the column density ratios
$N({\MgI})/N({\FeII})$ for each cloud and listed them in
Table~\ref{tab:acds}.  The $N({\MgI})/N({\FeII})$ data are plotted in
Figure~\ref{fig:rats}.  The horizontal bars of the data points give
the velocity interval of the $N(v)$ integrations and the vertical bars
give the uncertainties in the column density ratio.  Plotted in the
upper panels of Figure~\ref{fig:rats}, and aligned in velocity space
for ease of inspection, are the {\FeII} and {\MgI} $N(v)$ profiles.

As seen in Figure~\ref{fig:rats}, the ratios for all clouds in both
systems A and B are consistent with $\log N({\MgI})/N({\FeII}) \simeq
-2$.  This result suggests a high level of uniformity in both the
{\it gas--phase\/} [Mg/Fe] abundance patterns\footnote{We use the notation
$[X/Y]= \log (X/Y) - \log (X/Y)_{\odot}$ throughout this paper.} and
ionization conditions across velocity space.

The VP subcomponents describing
these dominant ``clouds'' also have ratios relatively consistent with
these findings, with $-2.02\pm0.04$ in system A and $-2.1\pm0.3$ in
system B.

Integration of $N(v)$ across the full profiles yields $\log
N({\MgI})/N({\FeII}) = -2.03_{-0.08}^{+0.07}$ for system A and
$-1.96_{-0.15}^{+0.16}$ for system B.  These quantities are shown as
grey shaded regions on Figure~\ref{fig:rats}.  The individual clouds
have ratios that are consistent with that of the full system.  This
level of uniformity with velocity is consistent with that reported by
\citet{prochaska02} for a sample of 13 higher redshift DLAs.  Worth
noting is that these data also provide direct evidence of {\it spatial
uniformity\/} on the scale of $135h^{-1}_{75}$ pc in a DLA.

\section{Nature of the Absorbers}
\label{sec:results}

\subsection{Photoionization Modeling}

Inferring the physical conditions in the clouds is model dependent.
Here, we assume photoionization equilibrium.  Using the measured
column densities as constraints on the models, we derive the cloud
metallicities, densities, sizes, and masses.

We used Cloudy \citep{ferland} to construct grids of model clouds.  We
assume that the $N({\HI})$ in system A clouds is $\log N =
20.3$~{\cmsq} and in system B clouds is $\log N = 19.9$~{\cmsq}
\citep{zuo97}.  Since the measured $N({\HI})$ is actually the sum of
individual clouds in a system, we note that some inferred properties
(sizes and masses) for the individual clouds will be overestimated.

Each cloud is modeled as a constant density plane--parallel ``slab''
with ionizing radiation incident on one face.  For the ionizing flux,
we used the $z=1.4$ ultraviolet background (UVB) spectrum of
\citet{handm96} and \citet{madau99}.  The separation velocity of the
DLA and the quasar is $\simeq 2500$~{\kms}.  Thus, it could be argued
that the absorber is not an intervening system, but is associated with
the quasar itself.  If so, the quasar flux, and not the UVB flux,
would dominate the photoionization of the gas.

One indicator of associated absorption is the presence of partial
covering \citep{barlow97,hamman97,ganguly99}.  None of the low
ionization transitions we observed show signs of partial covering.
That is to say, that the fully saturated regions of the {\MgIIdblt}
doublet members are black (statistically consistent with zero flux) in
their cores.  Also, at the same velocity intervals where {\FeII}
$\lambda 2383$ and $\lambda 2600$ are also fully saturated, their
absorption is also black.  As such, we assume that the DLA is
intervening.  If the absorber is intervening, then the cosmological
separation ($\Delta z = 0.02$) from the quasar is $\sim
15h^{-1}_{75}$~Mpc.  It has been argued that, at this distance, the
quasar flux does not affect the ionization balance in the absorber
\citep{michal97}.

At $h\nu = 13$~eV, the flux normalization is $\nu F_{\nu} = 1.26
\times 10^{-5}$ erg~s$^{-1}$~{\cmsq}.  We output the column densities
for selected ionic species (esp.\ {\MgII}, {\MgI}, and {\FeII}) as a
function of ionization parameter, $U=n_{\gamma}/n_{\rm H}$.  This
quantity is the ratio of the number density of photons capable of
ionizing hydrogen to the number density of hydrogen atoms.  We
explored the range $-8 \leq \log U \leq -1$ in 0.1 dex intervals.

For these large $N({\HI})$ values, the model clouds are optically
thick and have an extended neutral layer.  As such, the {\it
relative\/} dependence of the column densities with ionization
parameter are indistinguishable between the $\log N = 20.3$~{\cmsq}
and $\log N = 19.9$~{\cmsq} cloud models.

\subsection{Metallicity}

In Table~\ref{tab:ewlims}, we list the $3~\sigma$ rest--frame
equivalent widths of selected transitions.  
We infer upper limits on the cloud metallicities using the {\ZnII}
$\lambda 2026$ transition, which only weakly depletes onto dust and is
known to trace Fe--group elements \citep[e.g.,][]{savage96,jtl96}.  No
{\ZnII} was detected in our spectra; the $3~\sigma$ equivalent width
limit was $13$~m{\AA} in system A and $15$~m{\AA} in system B.
Without detections of {\CrII} and {\ZnII} we cannot directly estimate
the effects of dust in these systems \citep[see][]{zuo97}.

For these equivalent widths, {\ZnII} $\lambda 2026$ is effectively on
the linear portion of the curve of growth.  We obtain an upper limit
of $\log N({\ZnII}) < 11.9$~{\cmsq} for system A and $\log N({\ZnII})
< 12.0$~{\cmsq}. In the cloud models $N({\ZnII})$ is affectively
independent of ionization parameter for $\log U < -3.5$ and decreases
by only 0.5~dex from $\log U = -3.5$ to $-1$.  For $\log U < -3.5$, we
obtain an upper limit of [Zn/H]$ < -1$ for system A and [Zn/H]$ <
-0.5$ for system B.

These values are corroborated by upper limits based upon the weaker
{\ZnII} $\lambda 2063$ transition.  Our HIRES/Keck spectra provide
only slight improvements over the limits (also $3~\sigma$) of [Zn/H]$
< -0.75$ for system A and [Zn/H]$ < -0.31$ for system B reported by
\citet{pettini99}.

\subsection{Ionization Parameter: A Matter of Dust}

\begin{figure*}[p]
\plotone{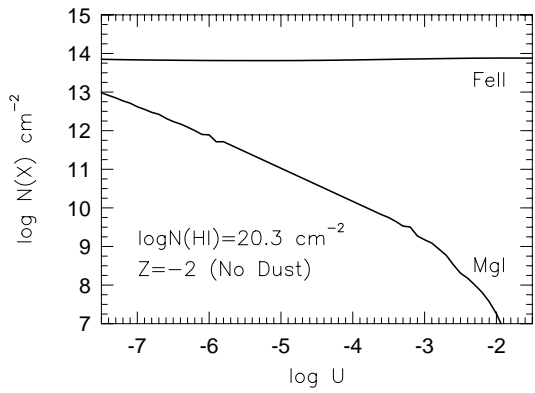}
\figurenum{5}
\caption
{A photoionization equilibrium Cloudy model of a plane--parallel cloud
with flux illuminated on one face.  The cloud has $\log N ({\HI}) =
20.3$~{\cmsq}, a metallicity of $[Z/Z_{\odot}] = -2$ and a solar
abundance pattern (i.e., no dust depletion).  Shown are $N({\FeII})$
and $N({\MgI})$ as a function of ionization parameter (see text).
\label{fig:cloudy}}
\end{figure*}

In the model clouds, $N({\MgII})$ and $N({\FeII})$ are virtually
independent of $U$, whereas $N({\MgI})$ decreases about 0.8~dex for
every 1~dex increase in $U$.  The $N({\MgI})/N({\MgII})$ ratio could
provide an estimate of the cloud ionization parameters; however, the
{\MgII} profiles are severely saturated.  Since the $N(v)$ profiles
for {\FeII} are well defined at all velocities, the
$N({\MgI})/N({\FeII})$ ratio, in principle, holds potential for
inferring the ionization parameter.  In Figure~\ref{fig:cloudy}, we
show the {\FeII} and {\MgI} column densities as a function of $\log U$
for one of our Cloudy models.  This model cloud has $\log N({\HI}) =
20.3$~{\cmsq} and a solar abundance pattern.  For presentation
purposed, we have scaled the metallicity to $[Z/Z_{\odot}] = -2$ to
match the integrated $N$ for system A presented in
Table~\ref{tab:acds}.  The logarithm of the ratio
$N({\MgI})/N({\FeII})$ decreases from $\simeq -0.5$ at $\log U = -8$
to $\simeq -4.5$ at $\log U = -3$, after which it begins to decrease
rapidly.  For a solar abundance pattern and no dust depletion, the
cloud models for both systems have $\log N({\MgI})/N({\FeII}) \simeq
-2$ for $\log U \simeq -6$.

The fact of the matter, however, is that $N({\MgI})/N({\FeII})$ is
directly proportional to the {\it gas--phase\/} [Mg/Fe] abundance
ratio, which can be strongly affected by the dust depletion factors
for Mg and for Fe.  The study of the effects of dust in DLAs is one of
central importance \citep[e.g.,][and references therein]{ellison01}.

\citet{zuo97} report a dust to gas ratio of $\sim 0.6$ for the
$z=1.3911$ DLA studied here.  However, we note that there are several
assumptions invoked for their calculations, including extrapolation of
the Galactic extinction law and no time delay or variability between
the lensed images.  The latter is certainly not strictly true
\citep[$\Delta \tau \simeq 420$ days, ][]{kundic97,haarsma99}, and the
former may not apply to DLAs \citep[e.g.,][]{pettini97,prochaska02}
even if the extrapolation accurately represents the Galactic law.
Even without a definitive estimate of the dust content in this DLA, we
can explore how dust depletion might effect the inferred cloud
properties.

In the Galaxy, dust depletion factors for Mg range from $\delta =
-0.3$ in warm ``halo'' gas to $\delta = -1.6$ in cool ``disk'' gas and
for Fe range from $\delta = -0.6$ to $-2.3$
\citep{jtl96,savage96,welty99,welty01}.  In order to allow for the
observed range in [Mg/Fe] due to dust depletion, we also consider
cloud models with warm halo and cool disk dust depletion factors.
These ranges also encompass the observed range of $\alpha$--group to
Fe--group variations observed in the photosphere of Galactic stars
\citep{mcwilliam95,jtl96,johnson02}, globular cluster stars
\citep{smith00,stephens02}, LMC/SMC stars \citep{venn98}, and dwarf
galaxies \citep{shetrone01}.

For the warm halo pattern we find the lower
limit of $\log U \simeq -6.2$ and for the cold disk pattern we find
the upper limit of $\log U \simeq -5.1$; we have,
\begin{equation}
-6.2 \leq \log U \leq -5.1
\end{equation}
for both the system A and system B clouds, where the no--dust, solar
abundance pattern value is $\log U = -6.0$.
 
We adopt this range in $\log U$ as an estimate of the inferred
ionization condition in the clouds.  We note, however, there is
growing observational evidence that dust depletion in DLAs may not
be significant and may be fairly uniform both from system to system
\citep{pettini99,ellison01} and from velocity component to velocity
component within a system \citep{prochaska02}.  However, we note that
there are examples of possible intrinsic abundance variations of
[Mn/Fe] in DLAs \citep{ledoux02} and of strong dust depletion in DLAs
selected because they are strong molecular hydrogen absorbers
\citep[e.g.,][]{petitjean02}.

The statistical data supporting low dust content in DLAs would suggest
that the above range of ionization parameters serves as a somewhat
conservative approach for bracketing the inferred ionization
parameter.  The largest uncertainty lurks in the original assumption
that the clouds are in photoionization equilibrium.

\subsection{Densities, Sizes, and Masses}

For the UVB spectrum normalized at $z=1.4$, the relationship between
the hydrogen number density of the clouds and the ionization parameter
is
\begin{equation}
\log n_{\rm H} = -4.9 - \log U~~ {\rm cm}^{-3},
\end{equation}
for constant density cloud models.  For the inferred range of
ionization parameters, we find
\begin{equation}
 3 \leq  n_{{\rm H},A} \leq 20~~ {\rm cm}^{-3},
\quad {\rm and}  \quad
 2 \leq  n_{{\rm H},B} \leq 15~~ {\rm cm}^{-3},
\label{eq:Urange}
\end{equation}
for the system A clouds and system B clouds, respectively.  These
densities fall within the lower density range for cold clouds in the
Galactic interstellar medium \citep{spitzer85,savage96}.  The cloud
line--of--sight physical extent, $D = N({\rm H})/n_{\rm H}$, is then,
\begin{equation}
\log D = \log N({\rm H}) + \log  U - 13.7~~ {\rm pc},
\end{equation}
where $N({\rm H})$ is the total hydrogen column density.  For the
inferred range of ionization conditions, the ionization fraction of
hydrogen is negligible so that the approximation $N({\rm H}) =
N({\HI})$ holds\footnote{The adopted Cloudy models yield hydrogen
ionization fractions slightly smaller than the value 0.05 reported by
Chartas et~al.\ (1995).}.  We obtain,
\begin{equation}
 5 \leq  D_{A} \leq 25 ~~ {\rm pc},
\quad {\rm and}  \quad
 1 \leq  D_{B} \leq 15 ~~ {\rm pc},
\label{eq:Drange}
\end{equation}  
for the system A and system B clouds, respectively.  To the extent
that the photoionization modeling has provided a reasonable
approximation of the cloud physical conditions, we can infer that the
cloud sizes are roughly a factor of ten smaller than the
sightline separation of the lensed quasar.

We can estimate the cloud mass under the assumption of spherical
symmetry, which gives $M \propto m_{\rm H}n_{\rm H} (D/2)^{3}$, where
$m_{\rm H}$ is the mass of hydrogen.  In terms of our
parameterizations, the cloud mass is,

\begin{equation}
\log M =  3 \log N_{\rm H} + 2\log U -47.7 ~~ {\rm M}_{\odot},
\end{equation}
which alternatively can be written as 
\begin{equation}
M \simeq 2 N_{20}^{3} U_{-6}^{2} ~~ {\rm M}_{\odot}, 
\end{equation}
where $N_{20}$ is the total hydrogen column density of the cloud in
units of $10^{20}$~{\cmsq}, and $U_{-6}$ is the ionization parameter
in units of $10^{-6}$.  We find approximate masses of
\begin{equation}
10 \leq M_{A}/M_{\odot} \leq 1000 
\quad {\rm and} \quad
0.5 \leq M_{B}/M_{\odot} \leq 60 ,
\end{equation}
for the system A and system B clouds, respectively, for the range of
ionization parameters given in Equation~\ref{eq:Urange}.

If we instead assume that the clouds are cylindrical ``slabs'' with
``height'' $D = N({\rm H})/n_{\rm H}$ and ``radius'' $R$, we have $M
\propto m_{\rm H}n_{\rm H}DR^{2}$, which can be simplified to
\begin{equation}
M \simeq 25 N_{20} R_{100}^{2} ~~ {\rm M}_{\odot} ,
\end{equation}
where $R_{100}$ is the cylinder radius in units of 100~pc.  Note that
the mass for this geometry is independent of the cloud density,
$n_{\rm H}$, and therefore ionization parameter, $U$.  Given that we
cannot track the individual clouds from sightline A to sightline B, a
conservative upper limit of $R_{100} = 2h_{75}^{-1}$ can be applied.
With this upper limit, we obtain
\begin{equation}
M_{A} \leq 400h_{75}^{-2}~~ {\rm M}_{\odot} 
\quad {\rm and} \quad
M_{B} \leq 160h_{75}^{-2}~~ {\rm M}_{\odot} ,
\end{equation}
for upper limits on the system A and system B cloud masses.

\subsection{Caveats}

We examined the effects on the above inferred cloud properties of
different shapes of the ionizing spectrum.  We examined a grid of
Cloudy models in which O stars and B stars contributed to and/or
dominated over the UVB.  We followed the formalism of \citet{cvc02}
and \citet{cl98} (esp., see Figures 12 and 13 of \citet{cvc02}).  We
find that the contribution of O and B stars makes virtually no
difference in the inferred cloud properties.  This is due to the high
level of self--shielding in the clouds.  Thus, we find that the
inferred cloud properties are robust under the assumption of
photoionization equilibrium.

It is important to point out that the above inferred sizes and masses
are overestimates within the formalism we have utilized.  Each
individual cloud was assumed to have identical $N({\HI})$, and the
value applied was the {\it total\/} $N({\HI})$ for each respective
system.  That is, all system A clouds were assumed to have $\log
N({\HI}) = 20.3$~{\cmsq} and all system B clouds were assumed to have
$\log N({\HI}) = 19.9$~{\cmsq}.  {\it Therefore, the inferred sizes
and masses for these clouds are smaller than we have quoted here.}

\section{Discussion}
\label{sec:discuss}

As first inferred from the FOS/{\it HST\/} spectra and confirmed here,
the absorber appears to have a projected transverse physical extent no
greater than $\simeq 135h^{-1}_{75}$~pc and seems to be enshrouded by
an optically thick gaseous complex \citep{michal97}.  We have resolved
the kinematics of the low ionization gas in these systems and find the
total velocity spread to be $\sim 300$~{\kms}, based upon the
{\MgIIdblt} doublet.

We have defined six ``clouds'' for system A and seven clouds for
system B.  The data and photoionization models suggest a picture in
which the sightline physical extent (sizes) of the clouds in this DLA
are of the order 10~pc.  The inferred densities are $\sim 10$~{\cc},
and the masses are no greater than $\sim 400h_{75}^{-2}$~M$_{\odot}$.
The temperatures of the cloud models, if they are taken at face value,
have a gradient with values that range from $T \simeq 2000$~K at the
cloud face to $T \simeq 500$~K at the cloud core.  To the extent that
the assumption of photoionization equilibrium is appropriate and has
been modeled accurately, we have found that the DLA clouds are similar
to low mass (10--100~M$_{\odot}$), low density (1--10~{\cc}), cold
($\sim 1000$~K) {\HI} clouds in the Milky Way
\citep{spitzer85,savage96}.

\subsection{Where lies the DLA?}

Directing our attention to the physical nature of DLAs, we focus on
two questions.  (1) Is the DLA region itself only one of these small
clouds, and therefore does it exhibit very narrow kinematics? (2) Is
the DLA region only a single phase of gas, or can it arise in a cold
phase and a warm phase \cite[e.g.,][]{lane00}?  The answer to these
questions will provide clues for interpreting the observed abundance
ratio and ionization uniformity between the clouds.

Based upon the cross--correlation (see Equation~\ref{eq:xcorr}) of the
system A and system B profiles, we find that the individual {\MgII}
and {\FeII} clouds are not identifiable between the two sightlines
separated by $135h^{-1}_{75}$~pc.  This implies that we do not track
the same clouds from sightline A to B, which in turn implies that they
are smaller than the line of sight separation.  \citet{rauch02-iv}
found a similar lack of a clear identification between clouds in an
optically thick {\MgII} absorber at $z=0.5656$ in the triple
sightlines of Q~$2237+0305$, which have separations of
135--200$h^{-1}_{75}$~pc.

However, the cross--correlation function has a single strong peak,
indicating that there is a single strongest component in each system;
system A has a strongest absorbing component at $v\simeq +20$~{\kms},
and system B has a dominant component at $v\simeq -10$~{\kms} (see
Figure~\ref{fig:aods}).  It is reasonable to argue that the DLA region
is physically associated with the sites of strongest absorption.  In
numerical simulations of hierarchical clustering, the strongest
absorption lines arise in the single most highly peaked baryon
overdensities \citep[e.g.][]{haehnelt96,rauch97}.

Because there are only low resolution spectra available for the neutral
hydrogen lines \citep[e.g.][]{turnshek93,michal97,zuo97,pettini99},
the measured $N({\HI})$ contains no direct kinematic information.
However, using the kinematics of the {\FeII} gas as a template for the
neutral hydrogen, we use the {\Lya}, {\Lyb}, {\Lyg}, {\Lyd}, and
{\Lye} profiles from the FOS spectrum in the {\it HST\/} archive to
test the hypothesis that the {\HI} arises in a single cloud.  While
the {\Lya} line constrains the damping wings, the higher
series lines provide constraints on the Doppler parameters.

We ran three simulations on the system A profiles: (1) a single cloud
at the redshift of the strongest {\FeII}--{\MgI} component; (2) six
clouds with equal $N({\HI})$ with velocities aligned with the
{\FeII}--{\MgI} components; and (3) six clouds with $N({\HI})$
proportional to $N({\FeII})$ and velocities aligned with the
{\FeII}--{\MgI} components.  For simulations ``2'' and ``3'', we
assumed all clouds had the same Doppler parameter.  The single cloud
model fit the data well; because of the higher order Lyman series, we
were able to put a limit of $50 \leq b({\HI}) \leq 60$~{\kms} on this
single cloud.  Although both multiple cloud simulations yielded good
fits to the higher order Lyman series, they totally failed to fit the
{\Lya} damping wings associated with $\log N({\HI}) = 20.3$~{\cmsq}.

Using the simulations as a guide, it is reasonable to assume that the
measured $N({\HI})$ for each sightline is the value associated with
the strongest component in systems A and B.  One scenario is that
these strongest clouds are physically linked.  If so, then the
velocity shear in the DLA is $\Delta v/\Delta r \simeq 30/135 =
0.22h_{75}$~{\kms}~pc$^{-1}$ and (for a planar geometry) the spatial
column density gradient is $\sim 6\times 10^{17}$~{\cmsq}~pc$^{-1}$.
It is more likely, however, that the clouds are individual parcels of
gas.  If so, then the A and B sightlines probe physically distinct
clouds with different $N({\HI})$ reflecting the structure of the
overall absorbing complex.

\subsection{A Two--Phase DLA?}

Using 21--cm observations of the $z=0.0912$ system toward the radio
quasar B~$0738+313$, \citet{lane00} reported a two phase DLA with
$T_{cold} \simeq 300$~K and $T_{warm} \simeq 5000$~K, where the warm
phase contributes roughly two thirds of the $N({\HI})$.  In the case
of Q~$0957+561$, VLBI mapping of the quasar
\citep[e.g.,][]{garrett94,campbell95} reveals that the extended radio
emission is probably not covered by the DLA.  \citet{kanekar02}
reported no detection of 21--cm absorption with an RMS of 2.4~mJy at a
resolution of 3.9~{\kms}.  These facts are consistent with the DLA
having a small covering factor.  It may not be possible to use 21--cm
absorption to constrain the phase structure of the DLA toward
Q~$0957+561$. 

However, based upon the VP fitting to the dominant components of the
{\MgI} profiles of the $z=1.3911$ DLA, and as can be seen in
Figures~\ref{fig:aods} and \ref{fig:rats}, the {\MgI} profiles have
broad wings.  The VP fits of these features yielded a single broad
component with $b=15.3\pm1.7$~{\kms} centered at $v=17.2$~{\kms} in
system A with $\log N({\FeII}) = 14.10\pm0.02$~{\cmsq} and $\log
N({\MgI}) = 12.08\pm0.05$~{\cmsq}.  In system B, three subcomponents
were found.  A narrow component with $b=5.6\pm1.8$~{\kms} centered at
$v=-3.3$~{\kms} with $\log N({\FeII}) = 13.92\pm0.04$~{\cmsq} and
$\log N({\MgI}) = 11.7\pm0.1$~{\cmsq}, a broader component with
$b=10.2\pm3.9$~{\kms} centered at $v=-14.9$~{\kms} with $\log
N({\FeII}) = 13.73\pm0.02$~{\cmsq} and $\log N({\MgI}) =
11.7\pm0.2$~{\cmsq}, and a third broader component with
$b=7.6\pm0.4$~{\kms} centered at $v=+14.0$~{\kms} with $\log
N({\FeII}) = 13.40\pm0.02$~{\cmsq} and $\log N({\MgI}) =
11.3\pm0.4$~{\cmsq}.

The broadness of the VP component in system A and the two adjacent VP
subcomponents in the wings of the system B component may be revealing a
warm phase analogous to that reported by \citet{lane00}.  If such
ionization structure were present, it would impact the inferences we
have on the cloud sizes and masses from our photoionization models.
We ran a fourth simulation on the Lyman series lines in which we model
a ``cold'' phase cloud and a ``warm'' phase cloud, each contributing
half the total $N({\HI})$ at the redshift of the strongest
{\FeII}--{\MgI} component.  

The maximum Doppler parameter the warm phase can have is $b=60$~{\kms}
as constrained by the higher order Lyman series lines.  We adopted
$b=55$~{\kms}.  By itself, this ``warm'' component, with $\log
N({\HI}) = 20.0$~{\cmsq}, cannot generate the {\Lya} damping wings
and also does not fit the {\Lyb} wings.  Testing a range of Doppler
parameters in the range $5 \leq b \leq 15$~{\kms} for the ``cold''
phase, we find that this component alone cannot account for a
substantial portion of the absorption in any of the Lyman series
lines.  However, together the two components provide a fit to the data
that is statistically identical to the single component model, with
$b<10$~{\kms} in the ``cold'' component yielding the best fit.

Therefore, we can confidently state that the {\HI} arises at a single
velocity, or velocity component; however, we cannot distinguish
between a single phase model and a ``cold+warm'' two--phase model.
We note however, that the inferred temperatures of our two--phase
model are higher than the temperatures reported by \citet{lane00} for
the $z=0.0912$ system toward B~$0738+313$.

It is difficult to incorporate a two--phase model into the direct
observation that the ion ratios are exceptionally uniform from
cloud--to--cloud across the full velocity extent of the profiles.  We
also find that this uniformity is spatial on the scale of
$135h^{-1}_{75}$~pc.  As stated above, this would suggest that the
gas--phase abundances of [Mg/Fe] are uniform both kinematically and
spatially vis--\`{a}--vis the ionization conditions.

Observational evidence is that Lyman limit systems are multiphase, and
they do not show kinematic uniformity.  Examination of
$N({\MgI})/N({\FeII})$ ratios in the Lyman limit systems studied by
\citet{cv01} reveals significant cloud to cloud variation with
velocity in the integrated apparent column densities (see their
Table~6). For example, the $z=1.0479$ {\MgII} system in PG~$0117+213$
has ratios $-1.03$ and $-1.55$ in its two kinematic subsystems.  The
$z=0.7729$ system in PG~$1248+401$ has a remarkable variation from
$-1.92$ to $-0.56$ in its two subsystems.  In addition, there are
examples of strong abundance and/or ionization variation with
kinematics \citep[e.g.,][]{ganguly98} and with spatial separation
\citep[e.g.,][]{rauch02-iv} in sub--Lyman limit systems.

\subsection{Systematic Kinematics?}

The kinematics of the {\MgI} profiles are reminiscent of the asymmetric
profiles studied by \citet{pw97,pw98}, who promoted the idea that DLA
gas arises in thick rotating disks of galaxies. However, in the case
of the $z=1.3911$ DLA toward Q~$0957+561$ the inferred size of the
individual cloud giving rise to this absorption is $\sim 10$~pc and
this precludes that this particular profile asymmetry is generated by
disk kinematics.

It is noteworthy that the velocity sheer of the strong absorbing
components is similar to the velocity sheer of the profile inversions
(regions where there is a paucity of gas; see first paragraph in
\S~\ref{sec:xcorr}).  This fact is indeed suggestive of some coherence
in the overall absorbing structure and dynamics, even if the
individual clouds cannot be directly traced across sightlines.  The
profile inversions may indicate physically separated gas parcels, as
suggested by the small cloud sizes we have derived.  If so, then the
common velocity sheer is remarkable; the multiple absorbing complexes
would be members of a generally extended, kinematically systematic
object \citep[a co--rotating halo?;][]{weisheit78,steidel02}.

It is clear from the higher ionization data \citep{michal97}, that
there is at least a low density, high ionization phase associated with
this system in addition to that giving rise to the {\MgII}, {\FeII},
and {\MgI}.  It is likely that this gas, especially the {\CIV}, is
more diffuse and extends more smoothly across the sightlines
\citep{rauch01-ii}, and is therefore not as directly coupled to the
low ionization systematic kinematics.

We examined if the velocity difference of $\simeq 30$~{\kms} between
the strongest {\MgI} clouds in the A and B profiles could arise due to
disk rotation.  Using the formalism of \citet{charlton95}, we modeled
the velocity sheer for a $\sim 200$~pc separation scale due to disk
kinematics for random sightline orientations.  We assumed a disk
circular velocities in the range of $150 \leq v_{c} \leq 250$~{\kms}.
We find that the sightlines cannot have velocity differences as large
as $30$~{\kms} but for highly contrived sightline--galaxy orientations
(the sightlines are too close together).

However, our models do not incorporate gas dispersion perpendicular
to the plane of the disk \citep[e.g.,][]{kinmods}.  Since such motion
is most definitely present in galactic disks, these models by no
means rule out the possibility that the absorbing material arises in a
disk; they do strongly suggest that the velocity sheer between
the two sightlines is not a result of simple disk kinematics.

\section{Conclusion}
\label{sec:conclusions}

We have observed the images of the lensed quasar Q~$0957+561$ A,B with
the HIRES/Keck--I instrument (resolution FWHM$~\simeq 6.6$~{\kms}).
We have presented an analysis of the {\MgII}, {\MgI}, and {\FeII}
absorption profiles from the $z=1.3911$ DLA system.  We adopted the
hydrogen column densities for systems A and B from \citet{zuo97},
which are $\log N({\HI}) = 20.3$ and 19.9~{\cmsq}, respectively.  The
line of sight separation is $\simeq 135h^{-1}_{75}$~pc at the
redshift of the absorber \citep{smette92}.

We converted the absorption profiles to their apparent optical depth
column density \citep[$N(v)$,][]{savage91}.  Based upon the location
of local minima in the $N(v)$ profiles for {\FeII}, we defined six
``clouds'' in system A and seven clouds in system B and integrated the
$N(v)$ to obtain the cloud column densities.  There is a ``dominant''
cloud in each line of sight.  It may be that these clouds contain the
bulk of the neutral hydrogen gas.  If the cloud geometry is planar and
extents across sightline B, then there is a neutral hydrogen column
density gradient of $9\times 10^{17}h_{75}$~{\cmsq}~pc$^{-1}$ and a
velocity sheer of $\simeq 0.2h_{75}$~{\kms}~pc$^{-1}$.

The clouds were assumed to be in photoionization equilibrium.  Using
Cloudy \citep{ferland}, we modeled the clouds as constant density,
plane--parallel ``slabs'' illuminated on one face by the ultraviolet
background ionizing spectrum.  We used the $N({\MgI})/N({\FeII})$
ratio in each cloud to constrain the ionization conditions.  Since
both Mg and Fe suffer dust depletion and originate predominantly in
separate nucleosynthetic environments, we bracketed the [Mg/Fe]
abundance pattern for the range of dust depletions seen in the Galaxy
and LMC/SMC and for the observed abundance patterns in the local
universe.

The observed $N({\MgI})/N({\FeII})$ ratio is remarkably uniform.  Not
only are the ratios consistent with $\log [N({\MgI})/N({\FeII})] = -2$
across the full velocity range in both systems, but they are also
consistent with this value spatially (in both sightlines).  This
resulted in very uniform cloud physical properties as inferred from
the photoionization modeling.  The ionization parameter of the clouds
is in the range $-6.2 \leq \log U \leq -5.1$.  This yielded clouds
with densities of $2 \leq n_{\rm H} \leq 20$~{\cc} and line of sight
physical extents of $1 \leq D \leq 25$~pc.  The inferred masses are
geometry dependent.  For spherical geometries the masses of the clouds
in system A are $10 \leq M/M_{\odot} \leq 1000$ and in system B are $1
\leq M/M_{\odot} \leq 60$.  For cylindrical geometries constrained by
the line--of--sight separation of less than 200~pc, the cloud masses
have upper limits of $400h_{75}^{-2}$~M$_{\odot}$ and
$160h_{75}^{-2}$~M$_{\odot}$ for systems A and B, respectively.  These
cloud properties are consistent with those for lower density, cold
clouds in the Galactic interstellar medium \citep{spitzer85,savage96}.

We focused our discussion on the physical nature of the DLA
``region'', the object that actually gives rise to the damped {\Lya}
absorption of $\log N({\HI}) = 20.3$~{\cmsq}.  Based upon simulations,
we favor a picture in which the DLA is a single cloud in the
multi--cloud profiles.  We cannot discern, however, if the DLA
comprises a ``cold'' single ionization phase, as suggested by our
photoionization models, or a ``cold+warm'' two--phase gas complex.  

If the DLA cloud is spherical in nature, then its size is on the order
of $\sim 10$~pc, and it is limited to one of the sightlines (A).  This
implies a covering factor of less than 0.1.  The other multiple gas
clouds in the proximity of this small DLA cloud would have to have
experienced the same sources of nucleosynthetic enrichment, be
optically thick in $N({\HI})$, and have similar dust contents.  This
implies that the material distributed in proximity to the DLA is well
mixed and ionized uniformly.  This is in stark contrast to the
significant variations seen in Lyman limit systems
\citep[e.g.,][]{cv01}, which are thought to arise in the outer disks
and halos of galaxies.  As such, we suggest that the low ionization
clouds accompanying DLAs are not arising in galactic halos.

Rather, we infer that DLAs arise in small gas--rich regions within
galaxies.  The data and models suggest that these regions are
complexes comprised of small, optically thick clouds similar to the
lowest mass, cold {\HI} clouds in the Galaxy.  Furthermore, the data
suggest that they are well mixed chemically and have similar
photoionization conditions.

\acknowledgements

Support for this work was provided by NASA (NAG 5--6399), and by the
NSF (AST 96--17185).  We thank Mike Keane for his assistance with the
observations.  We also thank Patrick Hall, Don Schneider, and an
anonymous referee for insightful comments that helped improved this
manuscript.




\begin{deluxetable}{lrrrrrrr}
\tablewidth{0pc}
\tablecolumns{8}
\tablecaption{Apparent Column Densities\tablenotemark{a}}
\tablehead
{
\multicolumn{8}{c}{System A}
 }
\startdata
Cld \#         & 1            & 2           & 3          & 4         & 5         & 6          &  \\
$(v_1,v_2)$    & $(-140,-88)$ & $(-88,-42)$ & $(-42,-3)$ & $(-3,35)$ & $(35,80)$ & $(80,124)$ & \\       
$N({\MgII})$   & $13.24_{-0.01}^{+0.02}$ & $>13.5$                 & $>13.6$                 & $>13.5$                 & $>13.6$                 & $>13.5$ & \\
$N({\FeII})$   & $13.23_{-0.02}^{+0.02}$ & $13.57_{-0.02}^{+0.02}$ & $13.70_{-0.02}^{+0.02}$ & $13.95_{-0.02}^{+0.03}$ & $13.69_{-0.02}^{+0.02}$ & $13.64_{-0.02}^{+0.03}$ & \\
$N({\MgI})$    & $ <10.6$                & $11.60_{-0.10}^{+0.09}$ & $11.54_{-0.11}^{+0.09}$ & $12.04_{-0.04}^{+0.04}$ & $11.65_{-0.09}^{+0.08}$ & $11.56_{-0.11}^{+0.09}$ & \\
{\MgI}/{\FeII} & $ <-2.6$                & $-1.97_{-0.10}^{+0.09}$ & $-2.16_{-0.11}^{+0.09}$ & $-1.91_{-0.04}^{+0.05}$ & $-2.04_{-0.09}^{+0.08}$ & $-2.08_{-0.11}^{+0.09}$ & \\
\cutinhead{System B}
Cld \#         & 1            & 2          & 3           & 4          & 5          & 6           & 7            \\
$(v_1,v_2)$    & $(-140,-90)$ & $(-90,56)$ & $(-48,-16)$ & $(-16,+5)$ & $(+5,+40)$ & $(+40,+85)$ & $(+85,+125)$ \\
$N({\MgII})$   & $>13.5$                 & $>13.4                $ & $>13.4                $ & $>13.3                $ & $>13.4                $ & $>13.6$                 & $>13.3                $ \\
$N({\FeII})$   & $13.47_{-0.02}^{+0.02}$ & $13.27_{-0.02}^{+0.02}$ & $13.40_{-0.02}^{+0.03}$ & $13.88_{-0.04}^{+0.05}$ & $13.48_{-0.02}^{+0.02}$ & $13.48_{-0.02}^{+0.02}$ & $13.07_{-0.02}^{+0.02}$ \\
$N({\MgI})$    & $11.14_{-0.39}^{+0.22}$ & $11.28_{-0.20}^{+0.15}$ & $11.47_{-0.12}^{+0.10}$ & $11.87_{-0.05}^{+0.05}$ & $11.65_{-0.08}^{+0.08}$ & $11.64_{-0.09}^{+0.09}$ & $11.31_{-0.19}^{+0.15}$ \\
{\MgI}/{\FeII} & $-2.33_{-0.39}^{+0.22}$ & $-1.99_{-0.20}^{+0.15}$ & $-1.93_{-0.12}^{+0.10}$ & $-2.01_{-0.06}^{+0.07}$ & $-1.83_{-0.08}^{+0.08}$ & $-1.84_{-0.09}^{+0.09}$ & $-1.76_{-0.19}^{+0.15}$ \\
\enddata
\tablenotetext{a}{All column densities are in units of atoms cm$^{-2}$.}
\label{tab:acds}
\end{deluxetable}

\begin{deluxetable}{lcc}
\tablewidth{0pc}
\tablecolumns{3}
\tablecaption{Equivalent Width Limits of Selected Transitions\tablenotemark{a}}
\tablehead
{
\colhead{Transition} &
\colhead{LOS A} &
\colhead{LOS B} \\
 &
\colhead{$W_{r,lim}$~{\AA}} &
\colhead{$W_{r,lim}$~{\AA}}
}
\startdata
{\MgI} 2026  & 0.012 & 0.016 \\
{\ZnII} 2026 & 0.013 & 0.015 \\
{\CrII} 2056 & 0.014 & 0.016 \\
{\CrII} 2062 & 0.011 & 0.014 \\
{\ZnII} 2063 & 0.011 & 0.013 \\
{\CrII} 2066 & 0.010 & 0.012 \\
{\FeII} 2250 & 0.011 & 0.013 \\
{\FeII} 2260 & 0.012 & 0.013 \\
{\NiI} 2312  & 0.010 & 0.013 \\
{\NiI} 2321  & 0.012 & 0.014 \\
{\FeI} 2484  & 0.010 & 0.013 \\
{\SiI} 2515  & 0.012 & 0.015 \\
{\FeI} 2524  & 0.010 & 0.014 \\
{\MnII} 2577 & 0.010 & 0.013 \\
{\MnII} 2594 & 0.027 & 0.017 \\
{\MnII} 2606 & 0.011 & 0.014 \\
\enddata
\tablenotetext{a}{All limits are rest--frame 3~$\sigma$.}
\label{tab:ewlims}
\end{deluxetable}


\end{document}